\documentclass[a4paper,11pt]{article}
\usepackage{pos}
\usepackage{tikz, float, adjustbox, hyperref, amsfonts}

\title{Axion QED as a Lattice Gauge Theory and Non-Invertible Symmetry}

\author*{Yamato Honda}
\author{Soma Onoda}
\author{Hiroshi Suzuki}

\affiliation{Department of Physics, Kyushu University,
744 Motooka, Nishi-ku, Fukuoka 819-0395, Japan}

\emailAdd{honda.yamato@phys.kyushu-u.ac.jp}
\emailAdd{onoda.soma@phys.kyushu-u.ac.jp}
\emailAdd{hsuzuki@phys.kyushu-u.ac.jp}

\abstract{
We investigate the non-invertible symmetry associated with chiral symmetry in axion quantum electrodynamics (QED) using the modified Villain formulation. In axion QED, it is known that naive magnetic objects such as 't~Hooft loops and axion strings lose their gauge invariance due to the violation of the Bianchi identity for the field strength of the photon or "field strength" of the axion. First, we construct the action of axion QED on the square lattice, which is more intricate than its counterpart in the continuum theory. We then observe the breaking of gauge invariance. Subsequently, we construct gauge-invariant magnetic objects by introducing new degrees of freedom localized at the positions of the magnetic objects. Furthermore, we explicitly compute the response of the magnetic objects under the action of the non-invertible symmetry operator constructed in Ref.~\cite{1}. In this analysis, we employ a method different from the so-called half-space gauging, which is the standard method to study non-invertible symmetries.
}

\FullConference{The 41st International Symposium on Lattice Field Theory (LATTICE2024)\\
 28 July - 3 August 2024\\
Liverpool, UK\\}

\begin{document}
\maketitle

\section{Introduction and summary}
Symmetry plays a key role in physics. In particular, global symmetries provide a framework for classifying phase structures and constraining possible infrared theories through obstructions arising from their gauging, known as 't~Hooft anomalies \cite{2}. On the other hand, in quantum electrodynamics (QED), it is known that the chiral symmetry, which is the global symmetry of this system, is broken at the quantum level due to the presence of the chiral anomaly \cite{3, 4}.

In recent years, attempts to generalize the concept of global symmetries, referred to as generalized symmetry \cite{5, 6, 7}, have been actively studied. Within this framework, the existence of symmetry is interpreted as the existence of symmetry operators. One of important contributions to this field is Refs.~\cite{8, 9}, which constructed symmetry operators that generate chiral transformations with rational angles in continuous QED. From the perspective of generalized symmetry, these results imply that chiral symmetry is partially preserved despite chiral anomaly. However, this symmetry is realized as a non-invertible symmetry \cite{10, 11, 12}, which has no inverse transformation. Indeed, it is known that non-invertible symmetry operators act non-invertibly on 't~Hooft loop operators.

In this paper, we investigate the non-invertible symmetry associated with chiral transformations in axion QED, the low-energy effective theory of QED, using lattice field theory (See Refs.~\cite{13, 14} for the continuum theory). Since lattice field theory is defined without any ambiguities, it allows us to analyze non-invertible symmetries without half-space gauging, which is the effective and standard method to discuss them. Furthermore, because all lattice fields are single-valued, lattice field theory is effective for treating singular configurations arising from magnetic objects such as 't~Hooft loops or axion strings. In particular, we employ modified Villain formulation \cite{15} on the square lattice (see Ref.~\cite{16} for a related work in $O(2)$ gauge theory). This formulation is well-suited to discuss non-invertible symmetries, as it describes the magnetic objects in terms of local variables.

The axion-photon coupling characterizing axion QED breaks gauge invariance at the locations of the magnetic objects (see Refs.~\cite{17, 18, 19, 20, 21, 13, 22} for the continuum theory). Equivalently, if we assume appropriate gauge transformations on the local variables describing the magnetic objects, the gauge invariance of the coupling can be restored. However, this results in the loss of gauge invariance for the naive magnetic objects. To make the magnetic objects gauge-invariant, we propose an introduction of additional degrees of freedom localized on them. Furthermore, we analyze the response of the gauge-invariant magnetic objects under the action of non-invertible symmetry operator constructed in Ref.~\cite{1}.
\section{Lattice formulation of axion QED}
\subsection{Modified Villain formulation}
In the modified Villain formulation \cite{15}, $U(1)$ gauge theory is construsted through $\mathbb{R}$ gauge theory. In $\mathbb{R}$ gauge theory, there is following 0-form $\mathbb{R}$ gauge symmetry $\mathbb{R}^{(0)}$ for the photon, denoted as $\mathbb{R}$-valued 1-cochain $a$,
\begin{align}
\mathbb{R}^{(0)}: \hspace{5mm} a \to a + \delta \lambda,
\end{align}
where $\lambda$ is $\mathbb{R}$-valued 0-cochain, and $\delta$ is the coboundary operator \cite{23, 24}, which is the lattce counterpart of the derivative $d$. In addition, there exists a 1-form $\mathbb{R}$ global symmetry.
Then, by gauging a subgroup $2 \pi \mathbb{Z}$ of the $\mathbb{R}$ global symmetry, we obtain $\mathbb{R} / 2 \pi  \mathbb{Z} \simeq U(1)$ gauge theory.
Specifically, we impose following 1-form $\mathbb{Z}$ gauge symmetry $\mathbb{Z}^{(1)}$ for the photon and so-called Villain field, represented by $\mathbb{Z}$-valued 2-cochain $z$, 
\begin{align}
\mathbb{Z}^{(1)}: \hspace{5mm} a \to a + 2 \pi m, \hspace{5mm} z \to z - \delta m,
\end{align}
where $m$ is $\mathbb{Z}$-valued 1-cochain.
The field strength of the photon, which is invariant under these gauge transformations, is given by
\begin{align}
f = \delta a + 2 \pi z.
\end{align}

We perform the same procedure for the axion, denoted as $\mathbb{R}$-valued 0-cochain $\phi$. 
The axion is characterized as a $2 \pi$ periodic scalar field, $\phi \sim \phi + 2 \pi$.
To realize this, we impose following 0-form $\mathbb{Z}$ gauge symmetry $\mathbb{Z}^{(0)}$ for the axion and the Villain field for the axion, denoted as $\mathbb{Z}$-valued 1-cochain $l$,
\begin{align}
\mathbb{Z}^{(0)}: \hspace{5mm} \phi \to \phi + 2 \pi k, \hspace{5mm} l \to l - \delta k,
\end{align}
where $k$ is $\mathbb{Z}$-valued 0-cochain.
"Field strength" of the axion, which is invariant under this gauge transformation, is given by
\begin{align}
\partial \phi := \delta \phi + 2 \pi l.
\end{align}
Therefore, the kinetic term $S_{\mbox{\footnotesize{kinetic}}}$ is expressed as
\footnote{Note that the convention in this paper is slightly different (though more familiar) from the one in the original works \cite{1, 26, 27}. The modified equations are all presented in this paper.}:
\begin{align}
S_{\mbox{\footnotesize{kinetic}}}
=
\sum _{\Gamma} \left(
\frac{1}{2 g _{0} ^{2}} f \cup \star f + \frac{\mu ^2}{2} \partial \phi \cup \star \partial \phi \right).
\end{align}
See Ref.~\cite{23, 24} for the cup product $\cup$, and Ref.~\cite{25} for the Hodge dual $\star$ defined on the original lattice instead of the dual lattice.

To preserve the topological charge on the lattice, we need to impose the Bianchi identities for the field strength of the photon and the axion by hand. Thus, we add the following to the action,
\begin{align}
S_{\mbox{\footnotesize{Villain}}}
&= \frac{1}{2 \pi} \sum _{\Gamma} ( i \tilde{a} \cup \delta f + i \delta \partial \phi \cup \chi ) \notag \\
&= \sum _{\Gamma} ( i \tilde{a} \cup \delta z + i \delta l \cup \chi ),
\label{1}
\end{align}
where the Lagrange multipliers $\tilde{a}$ and $\chi$ are respectively $\mathbb{R}$-valued 1-cochain, $\mathbb{R}$-valued 2-cochain. Indeed, integrating over $\tilde{a}$ and $\chi$ leads to the Bianchi identity for each,
\begin{align}
\delta z = 0, \hspace{5mm} \delta l = 0.
\end{align}
In fact, the integration of the Lagrange multipliers $\tilde{a}$ and $\chi$ corresponds to magnetic objects. The 't~Hooft loop $T _q (\gamma)$ and the axion string $S_{q'} (\sigma)$ are given by
\begin{align}
T _q (\gamma) 
= 
\exp \left( i q \sum_{\gamma} \tilde{a} \right)
, \hspace{5mm}
S_{q'} (\sigma) 
= 
\exp \left( i q' \sum_{\sigma} \chi \right),
\label{5}
\end{align}
for a closed loop $\gamma$, a two-dimensional closed surface $\sigma$, the magnetic charge of the 't~Hooft loop $q$ and the charge of the axion string $q'$.

Additionally, the insertion of the magnetic objects leads to the violation of the Bianchi identities only at their locations:
\begin{align}
\delta z = q \delta_3 [\gamma]
, \hspace{5mm}
\delta l = q' \delta_2 [\sigma],
\end{align}
where $\delta_3 [\gamma]$ and $\delta_2 [\sigma]$ are respectively represented as Poincar\'{e} dual of the corresponding manifold on the lattice.
\subsection{Construction of lattice action}
Let us construct the axion-photon coupling on the square lattice. The axion-photon coupling in the continuum theory is given by
\begin{align}
S ^{\mbox{\footnotesize{cont.}}} _{\mbox{\footnotesize{int.}}} = - \frac{i e^2}{8 \pi^2} \int_{\Gamma} \phi f \wedge f,
\end{align}
where the integer $e$ originates from the charge of the Dirac fermion present in QED, the high-energy theory of axion QED.
Here, we take $e$ to be even due to the requirement of the gauge invariance.
Since we aim to construct its lattice counterpart, we first attempt a naive discretization of the coupling in the continuum:
\begin{align}
- \frac{i e^2}{8 \pi^2} \int_{\Gamma} \phi f \wedge f
\Longrightarrow
- \frac{i e^2}{8 \pi^2} \sum_{\Gamma} \phi \cup f \cup f.
\end{align}
However, this discretization fails to yield the correct lattice action as it violates the $\mathbb{Z}^{(0)}$ gauge symmetry. Thus, we propose the following refinement of the lattice action:
\begin{align}
S_{\mbox{\footnotesize{int.}}}
=
- \frac{i e^2}{8 \pi^2} \sum_{\Gamma} \left\{ \phi \cup ( f \cup f + f \cup_1 \delta f) - 2 \pi l \cup (\mbox{CS}) \right\},
\end{align}
where (CS) is the Chern-Simons action on the lattice,
\begin{align}
 (\mbox{CS}) := a \cup f + 2 \pi z \cup a + 2 \pi a \cup_1 \delta z,
\end{align}
which is introduced by Ref.~\cite{24}.
See also Refs.~\cite{23, 24} for higher cup products such as $\cup_1$.
The construction of such a gauge-invariant lattice action can also be found in Ref.~\cite{28}, where it is applied to BF theory.

Despite the refinement of the lattice action, $S_{\mbox{\footnotesize{int.}}}$ transforms under the three gauge transformations as
\begin{align}
S_{\mbox{\footnotesize{int.}}} 
&\underset{\mathbb{Z}^{(0)}}{\longrightarrow} 
S_{\mbox{\footnotesize{int.}}} 
- i e^2 \sum_{\Gamma} k \cup a \cup \delta z,
\\
S _{\mbox{\footnotesize{int.}}} 
&\underset{\mathbb{Z}^{(1)}}{\longrightarrow} 
S _{\mbox{\footnotesize{int.}}} 
- \frac{i e^2}{2} \sum_{\Gamma} \delta l \cup m \cup a,
\\
S _{\mbox{\footnotesize{int.}}} 
\underset{\mathbb{R}^{(0)}}{\longrightarrow} 
S _{\mbox{\footnotesize{int.}}} + i e^2 \sum_{\Gamma} [- l \cup \lambda \cup &\delta z 
+ \delta l \cup \left\{ \lambda \cup \delta a + 2 \pi ( \lambda \cup z + z \cup \lambda + \lambda \cup_1 \delta z ) \right\} ].
\end{align}
Here, it is noteworthy that the violation of the gauge invariance takes the form of $* \cup \delta z$ and $\delta l \cup *$. Therefore, we can restore the gauge invariance by assigning the following appropriate gauge transformations to the Lagrange multipliers $\tilde{a}$ and $\chi$ in Eq.~$\eqref{1}$
\footnote{
The gauge transformations of the Lagrange multipliers are not essential. However, it is likely to make the discussion clearer.
}:
\begin{align}
\tilde{a} \underset{\mathbb{Z}^{(0)}}{\longrightarrow} \tilde{a} + e^2 &k \cup a
, \hspace{5mm}
\chi \underset{\mathbb{Z}^{(0)}}{\longrightarrow} \chi,
\label{2}
\\
\tilde{a} \underset{\mathbb{Z}^{(1)}}{\longrightarrow} \tilde{a} 
, \hspace{5mm}
&\chi \underset{\mathbb{Z}^{(1)}}{\longrightarrow} \chi + \frac{e^2}{2} m \cup a,
\label{3}
\\
\tilde{a} 
\underset{\mathbb{R}^{(0)}}{\longrightarrow} 
\tilde{a} + e^2 l \cup \lambda 
, \hspace{5mm}
\chi 
\underset{\mathbb{R}^{(0)}}{\longrightarrow} 
\chi - &\frac{e^2}{4 \pi} [ \lambda \cup \delta a + 2 \pi ( \lambda \cup z + z \cup 
\lambda + \lambda \cup_1 \delta z ) ].
\label{4}
\end{align}
Thus, the total action $S = S_{\mbox{\footnotesize{kinetic}}} + S_{\mbox{\footnotesize{Villain}}} + S_{\mbox{\footnotesize{int.}}}$ becomes completely gauge-invariant under the three gauge transformations.

\subsection{Construction of the 't Hooft loop}
Instead of preserving the gauge invariance of the action, the 't~Hooft loop and the axion string lose their gauge invariance (see Refs.~\cite{17, 18, 19, 20, 21, 13, 22} for the continuum theory). Indeed, the magnetic objects naively defined in Eq.~$\eqref{5}$ are no longer invariant under the gauge transformations.
't~Hooft loop and axion string are important physical observables to analyze the system, and it is necessary to recover the gauge invariance of the magnetic objects in some manner.

Here, we focus on the 't~Hooft loop $T _q (\gamma)$.
The method to restore the gauge invariance of the 't~Hooft loop is not unique.
Following Ref.~\cite{27}, we add new degrees of freedom that live only in $\gamma$.
Namely, we introduce $\mathbb{R}$-valued $0$-cochain $\varphi$, $\mathbb{Z}$-valued $1$-cochain $n$ and $\mathbb{R}$-valued $0$-cochain $\rho$.
Then, we assign the following gauge transformations to them: 
\begin{align}
\varphi &\underset{\mathbb{Z}^{(0)}}{\longrightarrow} \varphi + 2 \pi k,
\\
n &\underset{\mathbb{Z}^{(1)}}{\longrightarrow} n + e^2 q m,
\\
\rho &\underset{\mathbb{R}^{(0)}}{\longrightarrow} \rho + e^2 q \lambda.
\end{align}
For convenience, we define the gauge-invariant quantities as
\begin{align}
\partial \varphi &:= \delta \varphi + 2 \pi l,
\\
D_a \rho &:= \delta \rho + 2 \pi n -e^2 q a.
\end{align}
Consequently, we obtain the following gauge-invariant and genuine 't~Hooft loop operator:
\begin{align}
T _q (\gamma) 
&=
\exp \left( i q \sum _{\gamma} \tilde{a} \right)
\notag \\
&\times
 \int D \varphi D n D \rho \exp \left\{ \sum _{\gamma} \left(
  - \frac{l _{\varphi}}{2} \partial \varphi \cup \star \partial \varphi 
   - \frac{l _{\rho}}{2} D_a \rho \cup \star D_a \rho 
  + \frac{i}{2 \pi} \varphi \cup D_a \rho
   - i l \cup \rho \right) \right\},
\end{align}
where the coefficients of the kinetic terms, $l _{\varphi}$ and $l _{\rho}$, are parameters with the dimension of length.
\section{Action of non-invertible 0-form symmetry operator on charged operators}
Let us now turn to the 0-form non-invertible symmetry associated with chiral symmetry, the shift symmetry of the axion. From the action $S$, the equation of motion for the axion, 
\begin{align}
\delta \star j_5 = \frac{e^2}{8 \pi ^2} f \cup f \neq 0 ,
\end{align}
implies that the chiral current $j_5 := i \mu ^2 \partial \phi$ is not conserved due to the chiral anomaly. Thus, chiral symmetry seems to be absent in axion QED. However, chiral symmetry is realized as a non-invertible symmetry when the rotation angle is $\frac{2 \pi p}{N}$ for integers $p$ and $N$ that satisfy with $\mbox{gcd}(p e^2, N) = 1$.
Indeed, the following non-invertible 0-form symmetry operator can be constructed on the lattice (see Ref.~\cite{1}):
\begin{align}
U_{\frac{2 \pi p}{N}} (\mathcal{M}_3)  
    =
   \exp \left[ \frac{2 \pi p i}{N} \sum_{\mathcal{M}_3} \left\{
   \star j_5 - \frac{e^2}{8 \pi^2} (a \cup f + 2 \pi z \cup a) 
   \right\} \right] 
   \times \mathcal{Z}_{\mathcal{M}_3} [z].
\end{align}
Here, $\mathcal{Z}_{\mathcal{M}_3} [z]$ represents the lattice BF partition function on three-dimensional closed manifold $\mathcal{M}_3$, given by
\begin{align}
\mathcal{Z}_{\mathcal{M}_3} [z]
   =
   \frac{1}{N^s} \int Db Dc \exp \left[
   \frac{ip\pi e^2}{N}\sum_{\mathcal{M}_3}
   \left\{
   b(\delta c-z)-z\cup c
   \right\}
   \right],
\end{align}
where $b$ and $c$ are each $\mathbb{Z}_N$-valued $1$-cochain defined only on $\mathcal{M}_3$.

\begin{figure}
\centering
\begin{tikzpicture}[scale=0.4]
\draw[thick](7,3.9,0) circle(1.7 and 0.6);
\fill[red] (7,3.9,0) circle(1.7 and 0.6);
  \draw[thick] (0,5,0) ..controls (8,3,0) and (4,7,0).. (10,5,0) ..controls (10,4,-3) and (10,6,-6).. (10,5,-8) .. controls (4,6,-8) and (8,4,-8) .. (0,5,-8) ..controls (0,6,-6) and (0,4,-3).. (0,5,0);
  \node at (8,5,-4) {$\mathcal{M}_3'$};
  \draw[thick] (0,0,0) ..controls (8,-2,0) and (4,2,0).. (10,0,0) ..controls (10,-1,-3) and (10,1,-6).. (10,0,-8) .. controls (4,1,-8) and (8,-1,-8) .. (0,0,-8) ..controls (0,1,-6) and (0,-1,-3).. (0,0,0);
  \node at (8,0,-4) {$\mathcal{M}_3$};
  \node at (4.1,3.6,0) {$\mathcal{V}_4$};
  \node at (9.2,3.8,0) {$\gamma$};
  \node at (6.98,3.93,0) {$\mathcal{R}$};
  \draw (2,0) -- (3.5,1) -- (2.5,1) -- (2.5,4) -- (1.5,4) -- (1.5,1) -- (0.5,1) -- cycle;
  \fill[blue] (2,0) -- (3.5,1) -- (2.5,1) -- (2.5,4) -- (1.5,4) -- (1.5,1) -- (0.5,1) -- cycle;
\end{tikzpicture}
\caption{}
\label{10}
\end{figure}

We now examine the responses of charged operators under the action of the non-invertible symmetry operator that generates the chiral transformation (see Refs.~\cite{26, 27}).
Specifically, we consider deforming the symmetry operator from $\mathcal{M}'_3$ to $\mathcal{M}_3$ and sweeping the charged operators. In this case, there is a four-dimensional manifold $\mathcal{V}_4$, with boundary given by $\partial \mathcal{V}_4 = \mathcal{M}' _3 \cup (- \mathcal{M} _3)$ (see Fig.~\ref{10}).

In addition to the 't~Hooft loop $T _q (\gamma)$, there exist charged operators such as the axion operator $\Phi(x) = \exp (i \phi(x))$ for a point $x$ and the Wilson loop $W(C) = \exp (i \int _C a)$ for a closed loop $C$. Here, we assume the situation where $x, C, \gamma \subset \mathcal{V}_4$ The non-invertible symmetry operator acts invertibly, or equivalently, axially, on the axion operator: 
\begin{align}
\left\langle
U _{\frac{2 \pi p}{N}} (\mathcal{M}' _3)
\Phi(x)
\right\rangle
=
e^{\frac{2 \pi p i}{N}}
\left\langle
\Phi(x)
U _{\frac{2 \pi p}{N}} (\mathcal{M} _3)
\right\rangle.
\end{align}
Additionally, the non-invertible symmetry operator acts trivially on the Wilson loop:
\begin{align}
\left\langle
U _{\frac{2 \pi p}{N}} (\mathcal{M}' _3)
W(C)
\right\rangle
=
\left\langle
W(C)
U _{\frac{2 \pi p}{N}} (\mathcal{M} _3)
\right\rangle.
\end{align}

In contrast, the response of the 't~Hooft loop under the action of the non-invertible symmetry operator is more intricate. We first consider the case where the magnetic charge $q \not\in N \mathbb{Z}$ and $\gamma$ is non-contractable modulo $N$. In this case, there exists a two-dimensional closed manifold $\mathcal{M}_2$ that satisfies $\sum _{\mathcal{M}_2} z \not\in N\mathbb{Z}$ and $\mathcal{M}_2 \cap \gamma = \gamma$. As a result, the non-invertible symmetry operator vanishes the 't~Hooft loop due to the property of the lattice BF partition function:
\begin{align}
\left\langle
U _{\frac{2 \pi}{N}} (\mathcal{M}' _3)
T_q (\gamma)
\right\rangle
=
0.
\end{align}
Therefore, the non-invertible symmetry operator acts non-invertibly on the 't~Hooft loop. On the other hand, in all other cases, specifically when $q \not\in N \mathbb{Z}$ and $\gamma$ is contractable modulo $N$, or when $q \in N \mathbb{Z}$, the 't~Hooft loop acquires a surface of the field strength of the photon inside it: 
\begin{align}
\left\langle
U _{\frac{2 \pi p}{N}} (\mathcal{M}' _3)
T_q (\gamma)
\right\rangle
=
\left\langle
\exp \left( \frac{i p q e^2}{N} \sum _{\mathcal{R}} f \right)
T_q (\gamma)
U _{\frac{2 \pi p}{N}} (\mathcal{M} _3)
\right\rangle,
\end{align}
for a two-dimensional closed surface $\mathcal{R}$ that satisfies $\partial \mathcal{R} = \gamma$. These results are consistent with those reported in Ref.~\cite{13}. In particular, in the case where $q \in N \mathbb{Z}$, the surface of the field strength can be rewritten as a Wilson loop:
\begin{align}
\left\langle
U _{\frac{2 \pi p}{N}} (\mathcal{M}' _3)
T_q (\gamma)
\right\rangle
=
\left\langle
\exp \left( \frac{i p q e^2}{N} \sum _{\gamma} a \right)
T_q (\gamma)
U _{\frac{2 \pi p}{N}} (\mathcal{M} _3)
\right\rangle.
\end{align}
However, if we adopt a 't~Hooft loop that acquires a Wilson loop under the chiral transformation, such a Wilson loop does not appear. Therefore, the contribution of such a Wilson loop is not physically meaningful.

\section*{Acknowledgements}
 This work was partially supported by Kyushu University’s Innovator Fellowship Program (S.O.) and Japan Society for the Promotion of Science (JSPS) Grant-in-Aid for Scientific Research Grant Number JP23K03418 (H.S.).

\end{document}